\documentclass[prd,aps,twocolumn,a4paper,floatfix]{revtex4-1}

\usepackage{graphicx,psfrag}
\usepackage{mathrsfs}
\usepackage{amsmath,amsfonts,amssymb,amsthm}
\usepackage{hyperref}
\usepackage{url}
\usepackage[mathscr]{euscript}
\usepackage{bbm}
\usepackage{verbatim}
\usepackage{enumerate,cancel,wasysym}
\usepackage[utf8]{inputenc}

\usepackage{color}
\usepackage{float}
\usepackage{ulem}

\def\p{\partial}
\def\Lie{{\cal L}}

\def\non{\nonumber}
\def\ul{\underline}
\def\ov{\overset{\circ}}
\def\un{\underset{\circ}}

\def\gamN{{}^{\textrm{\tiny{(N)}}}\!\gamma}

\def\KN{{}^{\textrm{\tiny{(N)}}}\!K}

\def\gB{\mathbbmss{g}}
\def\gBi{(\mathbbmss{g}^{-1})}

\newtheorem{thm}{Theorem}
\newtheorem{prop}{Proposition}

\begin{document}

\title{Conformally flat slices of asymptotically flat spacetimes}

\author{Miguel Duarte$^{1,2}$}
\author{David Hilditch$^2$}

\affiliation{$^1$CAMGSD, Departamento de Matem\'atica, Instituto
  Superior T\'ecnico IST, Universidade de Lisboa UL, Avenida Rovisco Pais
  1, 1049 Lisboa, Portugal,
  $^2$CENTRA, Departamento de F\'isica, Instituto Superior
  T\'ecnico IST, Universidade de Lisboa UL, Avenida Rovisco Pais 1,
  1049 Lisboa, Portugal}

\begin{abstract}
For mathematical convenience initial data sets in numerical relativity
are often taken to be conformally flat. Employing the dual-foliation
formalism, we investigate the physical consequences of this
assumption. Working within a large class of asymptotically flat
spacetimes we show that the ADM linear momentum is governed by the
leading Lorentz part of a boost even in the presence of
supertranslation-like terms. Following up, we find that in spacetimes
that are asymptotically flat, and admit spatial slices with vanishing
linear momentum that are sufficiently close to conformal flatness, any
boosted slice can not be conformally flat. Consequently there are no
conformally flat boosted slices of the Schwarzschild spacetime. This
confirms the previously anticipated explanation for the presence of
junk-radiation in Brandt-Br\"ugmann puncture data.
\end{abstract}


\maketitle

\section{Introduction}\label{section:Introduction}

The construction of initial data for the Cauchy problem in General
Relativity (GR) relies on the one hand upon a suitable formulation of
the constraints, and on the other on a suitable choice for the given
data within this formulation. The former serves to provide a
theoretical framework in which data can be constructed, while the
latter encodes modeling choices of the physics under
consideration. Ideally this framework will allow for a straightforward
proof that data of physical interest exists, and for the required
given data to be easily interpreted. In numerical relativity the most
popular approach to solve the constraints is to make use of the
York-Lichnerowicz conformal transverse traceless
decomposition~\cite{Yor71,Yor72,Yor73} plus developments of the
approach culminating in the extended conformal thin-sandwich
equations~\cite{Ise08,BauCooSch98,Yor99,UryEri00,PfeYor02,BauSha10}. In
this setting the constraints become a coupled nonlinear elliptic
system. A particularly popular choice is the puncture
data~\cite{BraBru97}. Other strategies include
gluing~\cite{Cor00,CorSch03} and the reformulation of the constraints
as a hyperbolic evolution system~\cite{Rac15}.

An important aspect in formulating the constraints is the study of
exact solutions or those with special properties as this helps to
understand the physical nature of the constructed data. This is the
case for puncture data, which is in some sense inspired by the form of
Schwarzschild in isotropic coordinates. In fact, in all approaches
employing a free conformal metric, it often simplifies matters to make
that metric flat. Simplifying choices may however have unfortunate
physical consequences on the data being constructed. It is known, for
example, that the Kerr spacetime admits no spatial slice which is
conformally flat~\cite{Kro03,Kro04}. Therefore the use of this
restriction, even in the construction of a single spinning black hole,
must result in data which corresponds {\it not} to Kerr, but to some
physical deformation thereof. This deviation often appears as
high-frequency gravitational wave content, and is therefore referred
to as {\it junk-radiation}. Similar radiation is also observed in
evolutions of conformally flat initial data in which the black holes
have linear momentum. This feature becomes the crucial stumbling block
for highly boosted data~\cite{SpeCarPre08,SpeCarPre09}. With the
expectation that the restriction to conformal flatness was the cause
of this problem, several practical~\cite{ShiOkaYam08} and more
sophisticated~\cite{RucHeaLouZlo14,HeaRucLou15} cures have been
implemented. Another strategy is to try and account for the physical
effect of the junk-radiation. In the recent paper~\cite{HigKhaMcI19},
for example, a fitting method is used to do so.

The relationship between linear momentum, conformal flatness and
junk-radiation has notably been studied in the literature by
York~\cite{BowYor80,SmaYor78a,Yor80}, but usually with a fixed
background spacetime and a Taylor expansion in the boost. Here, to
avoid those simplifications, we employ the dual-foliation (DF)
formalism~\cite{HilRic13,Hil15,HilHarBug16,SchHilBug17,HilSch18,GasHil18}
and consider spacetimes which are asymptotically flat at
spatial-infinity. Properties of asymptotic charges, in particular of
the ADM 4-momentum are then examined under our definitions. We then
show that if there is a spatial slice with vanishing ADM-momentum
which is in some sense close to conformal flatness, then no slice
asymptotically related to the first by a boost near spatial-infinity
can be conformally flat. Morally this result can be summarized by
saying that {\it no slice of Schwarzschild with linear momentum is
  conformally flat} and therefore, in concordance with the expectation
mentioned above, conformal flatness {\it is} a cause of junk-radiation
in single black hole spacetimes with linear momentum. This is
presumably also true in a more general context.

We begin in section~\ref{section:As-Fl-ADM-4-Mom} with an overview of
the DF formalism and the various definitions and asymptotics that are
assumed afterwards. We demonstrate that the ADM 4-momentum is governed
by the leading Lorentz part of a boost even when supertranslation
terms are present. Section~\ref{section:Boosted-not-CF} contains the
main argument that, under refined assumptions on the asymptotics,
boosted slices can not be conformally flat. As a corollary we show
that axisymmetric slices of Kerr can not be conformally flat. We
conclude in section~\ref{section:Conclusions}. Geometric units are
used throughout.

\section{Asymptotic flatness and the ADM 4-momentum}
\label{section:As-Fl-ADM-4-Mom}

In this section we describe the DF formalism before giving a relevant
formulation of asymptotic-flatness at spatial-infinity. We then define
a change of coordinates that preserves this notion of asymptotic
flatness, and end by discussing the transformation of the ADM
energy-momentum under changes of coordinates that asymptote to
Poincar\'e transformations plus a supertranslation term near
spatial-infinity.

\subsection{DF formalism overview}

Given two families of observers, one associated with upper case
coordinates~$X^{\ul{\mu}}$, the other with the lower case~$x^{\mu}$,
spacetime will be described in two different but related ways. The DF
formalism~\cite{Hil15} provides a means to relate these two worldviews
from a~$3+1$ perspective. Throughout the paper, Latin
indices~$a,b,c,d,e$ will be abstract, underlined Greek indices denote
the components of tensors in the upper case coordinate tensor basis,
whereas plain Greek indices are used for the lower case
basis. Underlined and plain Latin indices ~$i,j,k,l$ stand for the
spatial components in the upper case and lower case bases
respectively.  The two time coordinates~$T$ and~$t$ provide, in
general, two distinct foliations of the spacetime, thus creating
different spatial tensors, spatial metrics, extrinsic curvatures and
so on. We denote with~$\gamN_{ab}$ the upper case spatial metric,
and~$\gamma_{ab}$ the lower case metric. The future pointing unit
normal vectors~$N^a$ and~$n^a$ of upper case and lower case foliations
are related by,
\begin{align}
N^a&=W(n^a+v^a)\,,
\end{align}
where we have defined the Lorentz factor~$W$ and lower case boost
vector~$v_a$,
\begin{align}
W& = -(N^an_a)\,,&\quad
v_a=\frac{1}{W}\perp^b\!\!{}_aN_b\,.
\end{align}
Here~$\perp^b\!\!{}_a$ is the projection operator on to the lower case
slice. Since the normal vectors have unit magnitude the Lorentz factor
and boost vector satisfy,
\begin{align}
W&=\frac{1}{\sqrt{1-v_iv^i}}\,,&\qquad W\geq1>\gamma^{ij}v_iv_j
\equiv v^2\,,
\end{align}
where~$\gamma^{ij}$ is the inverse lower case metric. Tensors
orthogonal on every slot to~$N^a$ and~$n^a$ are called upper case and
lower case respectively.  The~$3+1$ form of the spacetime
metric~$g_{ab}$ can be written as
\begin{align}
  ds^2&=
  (-A^2+B_{\ul{i}}B^{\ul{i}})dT^2+2B_{\ul{i}}dTdX^{\ul{i}}
  +\gamN_{\ul{ij}}dX^{\ul{i}}dX^{\ul{j}}\nonumber\\
  &=(-\alpha^2+\beta_i\beta^i)dt^2+2\beta_idtdx^i
  +\gamma_{ij}dx^idx^j\,,\label{eqn:4metric}
\end{align}
with standard definitions for the lapse and shift
variables. Subsequent definitions, such as that for the extrinsic
curvature of each foliation~$\KN_{ab}$ and~$K_{ab}$, follow the
standard lines. Their explicit relationship is given in~\cite{Hil15}.

The two tensor bases are of course related by the
Jacobian~$J^{\ul{\mu}}{}_\mu\equiv\p X^{\ul{\mu}}/\p x^{\mu}$, which
can we represented as,
\begin{align}
J&=\left(\begin{array}{cc}
A^{-1}W(\alpha-\beta^iv_i) 
& \alpha\,\pi^{\ul{i}}+\beta^i\phi^{\ul{i}}{}_i \\
-A^{-1}Wv_i 
& \phi^{\ul{i}}{}_i
\end{array}\right)\,,\label{eqn:Jacobian}
\end{align}
where~$\phi^{\ul{i}}{}_i\equiv J ^ {\ul{i}}{}_i$.

The projected upper case induced metric defined
by~$\mathbbmss{g}_{ab}=\gamma^c{}_a\gamma^d{}_b\gamN_{cd}$ is,
\begin{align}
\mathbbmss{g}_{ij}=\gamN_{ij}&=\gamma_{ij}+W^2v_iv_j\,.\label{eqn:GmN->gmn}
\end{align}
This object can be considered a metric on the lower case foliation and
it is called boost metric, with covariant derivative~$\mathbbmss{D}$
and connection~$\mathbbmss{G}$. The boost metric has inverse,
\begin{align}
(\mathbbmss{g}^{-1})^{ij}=\gamma^{ij}-v^iv^j\,.
\end{align}
For more details of the formalism we direct the reader
to~\cite{Hil15}.

\subsection{Asymptotic flatness}\label{section:Asymptotic_flatness}

Physically speaking, an asymptotically flat spacetime is characterized
by the requirement that the metric asymptotes to the Minkowski metric
sufficiently fast at large distances. A point well made
in~\cite{Ger77} is that no absolute preferred definition of asymptotic
flatness can be given or expected. Rather there is an interplay
between the field equations, the physics under consideration, and the
rate at which the metric becomes flat. Therefore several distinct
precise formulations of the concept have arisen. A key development in
these definitions has been the use of conformal
compactification~\cite{AshHan78}, which was used~\cite{Kro03,Kro04} in
the demonstration that there is no conformally flat slice of the Kerr
spacetime. We instead work with a more pedestrian definition, which is
motivated and stated in the following.

\paragraph*{Basic notion of asymptotic flatness:}
Consider a globally hyperbolic spacetime foliated by a family of
spacelike Cauchy hypersurfaces~$\Sigma_T$ and a boost-type
domain~$\Omega$ defined as,
\begin{align}
  \Omega:=\{R>R_0,|T|<qR+T_0\},
\end{align}
where~$R$ is a radial coordinate on that foliation, defined in the
standard way in terms of~$X^{\ul{i}}$, to be introduced momentarily,
and~$R_0$, $q>0$ and~$T_0$ are constants. The spacetime is said to be
{\it asymptotically flat} if there exists a preferred coordinate
system~$X^{\ul{\mu}}=(T,X,Y,Z)$, which will in general be highly
nonunique, with~$X^{\ul{i}}$ on~$\Sigma_T$, in which the
metric~$g_{ab}$ satisfies the following condition within~$\Omega$:
\begin{align}
	g_{\ul{\alpha\beta}}=\eta_{\ul{\alpha\beta}}+O_p(R^{-1})\,,\label{aflatness}
\end{align}
where~$\eta_{\ul{\alpha\beta}}$ is the Minkowski metric, $p\geqslant
1$ and~$O_p(R^{-m})$ means that its~$\p_{\ul{\alpha}}$ partial
derivatives of order~$n$ decay as $R^{-m-n}$ for all $n=0,\dots,p$.
Following a hint given in~\cite{Cho08}, we note that by combining the
boost theorem of~\cite{ChrMur81} with the improved Sobolev embedding
of~\cite{Bar86}, the present definition of asymptotic flatness can be
propagated from its natural restriction to initial data, in the vacuum
setting, inside a boost-type domain for any~$p\geq1$ provided that
suitable data, belonging to sufficiently high order weighted Sobolev
spaces, is given. Details can be found in
Appendix~\ref{section:App_Sob}.

\paragraph*{Permissible coordinate changes:}
If we are to restrict ourselves to the study of asymptotically flat
spacetimes, it is helpful to know which class of coordinate
transformations preserves the asymptotic form of the metric. In the
following we determine the form of these coordinate changes, which we
dub {\it permissible coordinate changes}. Intuitively, slices that we
can obtain from~$\Sigma_T$ through a permissible coordinate change are
called {\it permissible slices}. Let us assume momentarily that there
are two coordinate systems~$X^{\ul{\alpha}}$ and~$x^{\alpha}$ in which
the metric, close to spatial infinity, takes the
form~\eqref{aflatness}. Then we have,
\begin{align}
  &g_{\alpha\beta}=g_{\ul{\alpha\beta}}J^{\ul{\alpha}}\,_{\alpha}J^{\ul{\beta}}\,_{\beta}
  \nonumber\\
  \Rightarrow &\eta_{\alpha\beta}+O_p(r^{-1})
  =\eta_{\ul{\alpha\beta}}J^{\ul{\alpha}}\,_{\alpha}J^{\ul{\beta}}\,_{\beta}+O_p(R^{-1})
  \,.\label{metricupperlower}
\end{align}
In order to retrieve any information about the Jacobian, we have to
know how the two error terms relate to one another. For that we assume
that the upper case coordinate system can be expanded in powers
of~$r^{-1}$ in the following way:
\begin{align}
  X^{\ul{\alpha}}=X^{\ul{\alpha}}{}_0(t,\theta,\phi)r+X^{\ul{\alpha}}{}_1(t,\theta,\phi)
  + O_{p+1}(r^{-1})\,,\label{ansatzcoord}
\end{align}
where~$t:=x^0$ and~$\theta$ and~$\phi$ are the standard polar and
azimuthal angles associated with~$x^\alpha$. We can then write the
upper case radial coordinate near spatial infinity as,
\begin{align}
  R^2:=\delta_{\ul{ij}}X^{\ul{i}}{}X^{\ul{j}}{}
  =\delta_{\ul{ij}}X^{\ul{i}}{}_0X^{\ul{j}}{}_0r^2+O_{p+1}(r)\,,\label{Rr1}
\end{align}
which implies an equivalence of orders,
\begin{align}
O_p(r^{-q})=O_p(R^{-q})\,,\label{Rr2}
\end{align}
and we can conclude from~\eqref{metricupperlower} that the Jacobian
must have a leading Lorentz term,
\begin{align}
J^{\ul{\alpha}}\,_{\alpha}=\Lambda^{\ul{\alpha}}\,_{\alpha}+O_p(r^{-1})\,, \label{Jac1}
\end{align}
where~$\Lambda^{\ul{\alpha}}{}_\alpha$ is the standard Lorentz
matrix. Using the fact that~$J^{\ul{\alpha}}\,_{\alpha}:=\p_\alpha
X^{\ul{\alpha}}$, we can differentiate~\eqref{ansatzcoord} with
respect to~$x^\alpha$ and use~\eqref{Jac1} to get four equations, one
for each derivative of~$X^{\ul{\alpha}}$. The~$\p_i X^{\ul{\alpha}}$
equations give,
\begin{align}
    X^{\ul{\alpha}}{}_0r = \Lambda^{\ul{\alpha}}\,_{i}x^i\,,\label{x0}
\end{align}
and the~$\p_t X^{\ul{\alpha}}$ equation yields,
\begin{align}
  \p_t X^{\ul{\alpha}}{}_1 = \Lambda^{\ul{\alpha}}\,_t \Rightarrow X^{\ul{\alpha}}{}_1
  =\Lambda^{\ul{\alpha}}\,_t t+c^{\ul{\alpha}}(\theta,\phi)\,,\label{x1}
\end{align}
where~$c^{\ul{\alpha}}$ are arbitrary functions of lower case
angles. Plugging~\eqref{x0} and~\eqref{x1} in~\eqref{ansatzcoord}, we
get,
\begin{align}
    X^{\ul{\alpha}} =\Lambda^{\ul{\alpha}}{}_\alpha x^\alpha
  +c^{\ul{\alpha}}(\theta,\phi) +
  O_{p+1}(r^{-1})\,.\label{eqn:coordtransf}
\end{align}
We conclude that any coordinate transformation that preserves the
asymptotic form of the metric must have this form, and this is the
class that we will use throughout this work. Note that the Poincar\'e
transformations are precisely the subset of this large class with
constant~$c^{\ul{\alpha}}$ and vanishing error terms. For completeness
we write here the explicit form of the Jacobian in our notation, as
well as that of its inverse, which will be useful throughout this
work:
\begin{align}
J&=\left(\begin{array}{cc}
\bar{W} & -\bar{W}\bar{v}_j\delta^{j\ul{i}}\\
-\bar{W}\bar{v}_i& \delta^{\ul{i}}_i+\frac{\bar{W}^2\bar{v}_i\bar{v}_j}
{\bar{W}+1}\delta^{j\ul{i}}
\end{array}\right)+\p c(\theta,\phi) + O_{p}(r^{-2}) \,,\non\\
J^{-1}&=\left(\begin{array}{cc}
\bar{W} & \bar{W}\bar{v}^{i}\\
\bar{W}\bar{v}_j\delta^j_{\ul{i}}& \delta_{\ul{i}}^i+\frac{\bar{W}^2
  \bar{v}^i\bar{v}_j}
{\bar{W}+1}\delta^{j}_{\ul{i}}
\end{array}\right)+\p C(\Theta,\Phi) + O_p(R^{-2}) \,,\label{jac2}
\end{align}
with~$\bar{W}=(1-\bar{v}_i\bar{v}^i)^{-1/2}$, $\bar{v}_i$ constant and
$\bar{v}^i:=\bar{v}_j\delta^{ij}$. Now we need to check that all
transformations in the class~\eqref{eqn:coordtransf} preserve the
asymptotic form of the metric. For that we assume the metric to behave
like~\eqref{aflatness} and the coordinate transformation to be of the
form~\eqref{eqn:coordtransf}, and compute the behavior
of~$g_{\alpha\beta}$ close to spatial infinity,
\begin{align}
g_{\alpha\beta}&=g_{\ul{\alpha\beta}}J^{\ul{\alpha}}\,_{\alpha}J^{\ul{\beta}}\,_{\beta}
=\eta_{\ul{\alpha\beta}}J^{\ul{\alpha}}\,_{\alpha}J^{\ul{\beta}}\,_{\beta}+O_p(R^{-1})
\nonumber\\
& = \eta_{\alpha\beta} +O_p(r^{-1})+O_p(R^{-1})= \eta_{\alpha\beta}
+O_p(r^{-1})\,,\non
\label{glowercaseAs}
\end{align}
where the last equality comes from~\eqref{Rr2}. This concludes the
proof that the largest class of coordinate transformations that
preserve the asymptotic behavior required by our definition of
asymptotic flatness~\eqref{aflatness} is given
by~\eqref{eqn:coordtransf}. We can then write the following result:
\begin{prop}
    Let $(M,g)$ be an asymptotically flat spacetime with preferred
    coordinates $X^{\ul{\alpha}}$. A coordinate transformation is
    permissible if and only if it is of the form,
\begin{align}
    X^{\ul{\alpha}} =\Lambda^{\ul{\alpha}}{}_\alpha x^\alpha
  +c^{\ul{\alpha}}(\theta,\phi) +
  O_{p+1}(r^{-1})\,.
\end{align}
\end{prop}
It is straightforward to see that the
inverse transformation takes the analogous form,
\begin{align}
  x^{\alpha} =(\Lambda^{-1})^{\alpha}{}_{\ul{\alpha}} X^{\ul{\alpha}}
  +C^{\alpha}(\theta,\phi) +
  O_{p+1}(R^{-1})\,,\label{eqn:inverse_coordtransf}
\end{align}
with the new quantities and~$O$ defined in the obvious manner,
and~$C^{\alpha}=-(\Lambda^{-1})^{\alpha}{}_{\ul{\alpha}}c^{\ul{\alpha}}$. Additionally,
the angular coordinates can be seen to satisfy,
\begin{align}
  \Theta&=\frac{\Lambda^Z{}_ix^i}{R} + O_{p+1}(r^{-1})\,,\nonumber\\
  \Phi&=\frac{\Lambda^Y{}_ix^i}{\Lambda^X{}_jx^j} + O_{p+1}(r^{-1})\,.
  \label{angles}
\end{align}
Note that the leading order terms in~$\Theta$ and~$\Phi$ depend only
on~$\theta$ and~$\phi$, which is why we can define~$C^{\alpha}$ in
terms of~$\theta$ and~$\phi$ in~\eqref{eqn:inverse_coordtransf}.

\paragraph*{The ADM~4-momentum:} The ADM~$4$-momentum is defined as,
\begin{align}
P^{\textrm{ADM}}_{\ul{\alpha}} := (-m,P_X,P_Y,P_Z)\,,\label{eqn:momtransf}
\end{align}
where~$m$ is the ADM mass and~$P_{\ul{i}}$ are the components of the
ADM linear momentum, given in terms of the intrinsic metric and
extrinsic curvatures by,
\begin{align}
&P^{\textrm{ADM}}_{T}=\frac{-1}{16\pi}\lim\limits_{R\rightarrow \infty}
\int_{S_R}(\p_{\ul{i}}\gamN_{\ul{ij}}-\p_{\ul{j}}\gamN_{\ul{ii}})\,dS_R^{\ul{j}}\,,\\
&P^{\textrm{ADM}}_{\ul{i}}=\frac{1}{8\pi}\lim\limits_{R\rightarrow \infty}
\int_{S_R}(\KN_{\ul{ij}}-\KN\gamN_{\ul{ij}})\,dS_R^{\ul{j}}\,.\label{eqn:ADMmom}
\end{align}
Here,~$S_R$ is a coordinate 2-sphere of radius~$R$. Our definition of
asymptotic flatness is sufficient for the ADM four-momentum to be well
defined. $P^{\textrm{ADM}}_{\ul{\alpha}}$ behaves as a~$4$-dimensional
linear form under coordinate
change~\eqref{eqn:coordtransf}~\cite{Gou07}. In fact, ADM
showed~\cite{ArnDesMis62} that a Poincar\'e transformation transforms
the $4$-momentum according to,
\begin{align}
P^{\textrm{ADM}}_{\ul{\alpha}}=\Lambda_{\ul{\alpha}}{}^\alpha
P^{\textrm{ADM}}_{\alpha}\,,\label{eqn:ADMmom-transformation}
\end{align}
where~$\Lambda_{\ul{\alpha}}{}^\alpha=\Lambda^{\ul{\beta}}{}_\beta
\eta_{\ul{\alpha\beta}}\eta^{\alpha\beta}$. For an introduction of the
ADM conserved quantities at spatial infinity based on
differentiability requirements for the Hamiltonian
see~\cite{BeigMurch}. Here a slightly more restrictive definition of
asymptotic flatness is used.

\subsection{Supertranslations do not affect the transformation
  of the ADM 4-momentum}

We work in this subsection along the lines of the discussion given
in~\cite{PiotrEner} (Section 1.2.3). The Einstein-Hilbert action
contains second derivatives of the metric, but one can remove a total
divergence from it, leaving the action with only first derivatives of
the metric. As the term removed is a total divergence, the field
equations are unchanged. This can be achieved by introducing a
background metric.

\begin{figure}[t!]
    \includegraphics[width=0.48\textwidth]{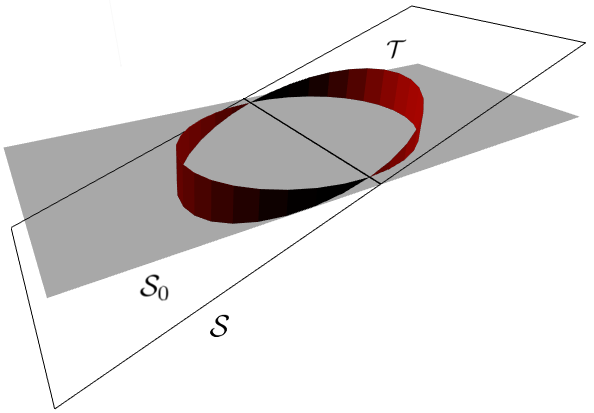}
    \caption{\label{Fig:T6} Depiction of the geometry involved in
      Theorem~\ref{Thm:T1}. The gray-shaded surface denotes the lowercase
      (boosted) spatial slice and the white the uppercase. To relate
      the asymptotic charges on each, the divergence law is applied on
      the red surface~$\mathcal{T}$.}
\end{figure}

Let~$\mathcal{S}_{0}$ be the spacelike hypersurface~$\{T=0\}\cap
\Omega$, where~$\Omega$ is a boost-type domain. On~$\mathcal{S}_{0}$,
we define the upper case background metric~$B_{ab}$ by requiring
that~$B_{\ul{\alpha\beta}}=\eta_{\ul{\alpha\beta}}$. Note that, due to
the Poincar\'e-invariance of the Minkowski metric, if we
took~$c^{\ul{\alpha}}$ in~\eqref{eqn:coordtransf} constant, there
would be no difference between~$B_{ab}$ and the analogously defined
lower case background metric~$b_{ab}$. In our case, that difference
takes the form,
\begin{align}
  B_{\alpha\beta}-b_{\alpha\beta}&=2\p_{(\alpha}[\Lambda^{\ul{\alpha}}{}_{\beta)}
    c^{\ul{\beta}}\eta_{\ul{\alpha\beta}}] +O_{p}(R^{-2})\nonumber\\
	&=:2\p_{(\alpha}\tilde{c}_{\beta)}+O_{p}(R^{-2})\,. \label{diffBb}
\end{align}
We define the following tensor,
\begin{align}
  \ov{\mathfrak{g}}\,^{ab}
  :=\frac{1}{16\pi}\frac{\sqrt{-g}}{\sqrt{-B}}\,g^{ab}\,,\label{upperdens}
\end{align}
where~$g$ and~$B$ are the determinants of~$g_{ab}$ and~$B_{ab}$,
respectively. As in~\cite{PiotrEner} (Section 1.2.3), given a vector
field~$X$, the Hamiltonian generating the flow of~$X$ can be written
as,
\begin{align}
  H(X,\mathcal{S}_{0})
  &:=\int_{\mathcal{S}_{0}}\big( p^a_{bc}\Lie_X\ov{\mathfrak{g}}{}^{bc}
  -X^aL\big)dS_a \nonumber\\
  &=\int_{\mathcal{S}_{0}}\ov{\nabla}_b \ov{\mathbb{U}}{}^{a b}dS_a\,,\non\\
  \ov{\mathbb{U}}{}^{ab}&:=\ov{\mathbb{U}}{}^{ab}{}_{\;c}X^c\,
  - 2\ov{\mathfrak{g}}{}^{d[a}\delta^{b]}_c \ov{\nabla}_d X^c\,,\label{Udef}\non\\
  \ov{\mathbb{U}}{}^{ab}{}_{\;c}&:=2 (\ov{\mathfrak{g}}{}^{-1})_{c d}
  \ov{\nabla}_e \left(\ov{\mathfrak{g}}\,^{d[b}
    \ov{\mathfrak{g}}\,^{a]e}\right)\,,
\end{align}
where~$\ov{\nabla}_a$ is the covariant derivative associated
with~$B_{ab}$, $L$ is the Lagrangian and~$p^a_{bc}$ is the momentum
canonically conjugate to~$\ov{\nabla}\ov{\mathfrak{g}}\,^{ab}$ defined
as,
\begin{align}
p^a_{bc}:=\frac{\p L}{\p (\ov{\nabla}_a\ov{\mathfrak{g}}\,^{bc})}\,.
\end{align}
The ADM 4-momentum on~$\mathcal{S}_{0}$ is then,
\begin{align}
P^{\textrm{ADM}}_{\ul{\alpha}}:=H(\p_{\ul{\alpha}},\mathcal{S}_{0})\,.\label{pdef}
\end{align}
By Chru\'sciel's Proposition 1.2.1 in~\cite{PiotrEner}, if the
metric~$g_{ab}$ satisfies our notion of asymptotic
flatness~\eqref{aflatness}, then the
integral~$H(\p_{\ul{\alpha}},\mathcal{S}_{0})$ converges. Now
let~$\mathcal{S}$ be the spacelike hypersurface~$\{t=0\}\cap \Omega$
which is related to~$\mathcal{S}_{0}$ by a change of coordinates of
the form~\eqref{eqn:coordtransf}. As we have shown that the boundary
conditions are preserved under~\eqref{eqn:coordtransf}, we immediately
get convergence of the integral $H(\p_{\ul{\alpha}},\mathcal{S})$.  In
order to show that~\eqref{eqn:ADMmom-transformation} is unchanged by
supertranslations, we define the following 3-dimensional region of
spacetime,
\begin{align}
\mathcal{T}=\{R=R_0,T>0,t<0\}\cup\{R=R_0,T<0,t>0\}\,,\nonumber
\end{align}
so that its boundary $\p\mathcal{T}$ consists of two 2-spheres of
radius~$R_0$, $\mathcal{S}_0\cap \{R=R_0\}$ and~$\mathcal{S}\cap
\{R=R_0\}$ respectively, as shown on Fig.~\ref{Fig:T6}.
%
%
We integrate~$\ov{\nabla}_b \ov{\mathbb{U}}{}^{a b}$
over~$\mathcal{T}$ and use Gauss's theorem to write,
\begin{align}
  \int_{\mathcal{S}\cap \{R=R_0\}} \ov{\mathbb{U}}{}^{ab}dS_{ab}
  = \int_{\mathcal{T}}\ov{\nabla}_b \ov{\mathbb{U}}{}^{a b}dS_a
  + \int_{\mathcal{S}_0\cap \{R=R_0\}} \ov{\mathbb{U}}{}^{ab}dS_{ab}.
  \label{eqn:Gauss}
\end{align}
From~\cite{PiotrEner} we can see that the integrand in the first term
on the right-hand side can be written as,
\begin{align}
16\pi\ov{\nabla}_b \ov{\mathbb{U}}{}^{a b}
= \sqrt{-B}(\mathbb{T}^a{}_bX^b + Q^a{}_bX^b
+Q^{ab}{}_c\ov{\nabla}_bX^c)\,,\nonumber
\end{align}
where~$\mathbb{T}^a{}_b$ is called canonical stress and is defined as,
\begin{align}
\mathbb{T}^a{}_b:=\frac{1}{8\pi}\frac{\sqrt{-g}}
{\sqrt{-B}}\left(R^a{}_b-\frac{1}{2}Rg^a{}_b\right)\,,
\end{align}
$Q^a{}_b$ is, to leading order, quadratic in~$\ov{\nabla}g_{ab}$
and~$Q^{ab}{}_c$ is bilinear in~$\ov{\nabla}g_{ab}$
and~$g_{ab}-B_{ab}$ with bounded coefficients. At this point we have
to make the additional assumption that the spacetime satisfies
Einstein's equations with a stress-energy tensor decaying as
$O_{0}(R^{-4})$. This implies that~$\mathbb{T}^a{}_b$ decays near
spatial infinity at least as,
\begin{equation}
    \mathbb{T}^{\ul{\alpha}}{}_{\ul{\beta}}=O_{0}(R^{-4})\,.\label{Tfalloff}
\end{equation}
This requirement is necessary because the integral on $\mathcal{T}$
involves integrating over two angular coordinates, yielding $R_0{}^2$
in the volume element, and one time coordinate, yielding $R_0$ as the
time interval grows with $R_0$. Then, as~$R_0\to \infty$, the first
term on the right-hand side of~\eqref{eqn:Gauss} is,
\begin{align}
\int_{\mathcal{T}}\ov{\nabla}_b \ov{\mathbb{U}}{}^{a b}dS_a =
O_{0}(R_0^{-1})\,,
\end{align}
and the remaining two terms give,
\begin{align}
\lim_{R_0\to\infty} \int_{\mathcal{S}\cap \{R=R_0\}} \ov{\mathbb{U}}{}^{ab}dS_{ab}
=H(\p_{\ul{\alpha}},\mathcal{S}_0)\,.\label{Gauss2}
\end{align}
We would like the
left-hand side to reduce to~$H(\p_{\ul{\alpha}},\mathcal{S})$. For
that, we have to rewrite the integrand in terms of the following
tensor,
\begin{align}
  (\un{\mathfrak{g}}^{-1})\,_{ab}:=
  \frac{1}{16\pi}\frac{\sqrt{-b}}{\sqrt{-g}}\,g_{ab}
  =\frac{\sqrt{-b}}{\sqrt{-B}}\,
  (\ov{\mathfrak{g}}{}^{-1})_{ab}\,,\label{lowerdens}
\end{align}
where~$b$ is the determinant of the lower case background
metric, and replace the covariant
derivative~$\ov{\nabla}_a$ with the one associated
to~$b_{ab}$, $\un{\nabla}{}_a$, by making use of the tensor,
\begin{align}
\mathbb{C}_{b}{}^{a}{}_{c}:=&\ov{\Gamma}_{b}{}^{a}{}_{c}
-\un{\Gamma}{}_{b}{}^{a}{}_{c}\nonumber\\
=&\frac{1}{2}(B^{-1})^{ad}(\un{\nabla}{}_b B_{cd}
+\un{\nabla}{}_c B_{bd}-\un{\nabla}{}_d B_{bc})\,, \label{dualChrist}
\end{align}
where~$\ov{\Gamma}_{b}{}^{a}{}_{c}$
and~$\un{\Gamma}{}_{b}{}^{a}{}_{c}$ are Levi-Civita connections
of~$B_{ab}$ and~$b_{ab}$, respectively.  In order to compute the ADM
4-momentum, the vector~$X$ is chosen to be $\p_{\ul{\alpha}}$ (see
\eqref{pdef}) and the second term on the definition
of~$\ov{\mathbb{U}}\,^{ab}$~\eqref{Udef} vanishes, so let us write,
\begin{align}
\ov{\mathbb{U}}\,^{ab}\,_c = &2\frac{\sqrt{-b}}{\sqrt{-B}}
\Bigl[\frac{1}{2}\un{\mathbb{U}}^{ab}\,_c+\frac{B}{b}\delta_c^{[b}g^{a]e}
\un{\nabla}{}_e\frac{b}{B}\nonumber\\
&+ \mathbb{C}_{e}{}^{e}{}_{f}\delta_c^{[b}g^{a]f}
+ \mathbb{C}_e{}^{[b}{}_cg^{a]e}
+ \mathbb{C}_e{}^{[a}{}_{f}\delta_c^{b]}g^{ef}\Bigr]\,.\label{Uchange}
\end{align}
The quotient of the determinants of the background metrics can be
computed using~\eqref{diffBb}:
\begin{align}
	\frac{b}{B}=1-2(b^{-1})^{ab}\un{\nabla}{}_a\tilde{c}_b +O_{p}(R^{-2})\,.
\end{align}
We must now define the future-pointing vectors normal
to~$\mathcal{S}_0$ and~$\mathcal{S}$, normalized with respect to the
background metrics. Respectively,
\begin{align}
  &\ov{n}_a:=-\ov{\alpha}\un{\nabla}{}_at\,, \quad
  \ov{\alpha}{}^{-2}:=-(B^{-1})^{bc}\un{\nabla}{}_bt
  \un{\nabla}\,_ct\,,\nonumber\\
  &\un{n}{}_a:=-\un{\nabla}{}_at\,.\label{alpha}
\end{align}	
We must also define the outward-pointing vectors normal
to~$\mathcal{S}_0\cap \{R=R_0\}$ and~$\mathcal{S}\cap\{R=R_0\}$,
\begin{align}	
\ov{s}_a&:=\ov{\nabla}_aR\,,\nonumber\\
\un{s}{}_a&:=\ov{\gamma}{}_a^b\ov{s}_b\,,\quad
\ov{\gamma}{}_a^b:=\delta_a^b
+ \ov{n}_a\ov{n}_c(B^{-1})^{bc}\,,\label{sup}
\end{align}
respectively, in order to build the integrand on the left-hand side
of~\eqref{Gauss2}:
\begin{align}
  &\ov{\mathbb{U}}\,^{ab}\,_c\ov{n}_a \ov{s}_b X^c
  = \un{\mathbb{U}}\,^{ab}\,_c\ov{n}_a \ov{s}_b X^c \nonumber\\
  & + 2\un{\nabla}{}_f\un{\nabla}{}_d\tilde{c}_e\Bigl[g^{fb}g^{e[d}
  - g^{fd}g^{e[b}\Bigr]\ov{n}_a \ov{s}_b X^{a]}
  + O_{p-1}(R^{-3})\,.\label{integrand1}
\end{align}
In order to understand the last non-error term in~\eqref{integrand1},
we have to perform the following simple computation,
\begin{align}
\ov{s}_{\ul{\alpha}}\un{\nabla}{}_{\ul{\delta}}
&\un{\nabla}^{\ul{\delta}}\tilde{c}_{\ul{\beta}}
- \ov{s}{}^{\ul{\delta}}\un{\nabla}{}_{\ul{\delta}}
\un{\nabla}{}_{\ul{\alpha}}\tilde{c}_{\ul{\beta}}
= \un{\nabla}{}_{\ul{\delta}}(\ov{s}_{\ul{\alpha}}
\un{\nabla}^{\ul{\delta}}\tilde{c}_{\ul{\beta}})
- \un{\nabla}{}_{\ul{\delta}}\ov{s}_{\ul{\alpha}}
\un{\nabla}^{\ul{\delta}}\tilde{c}_{\ul{\beta}} \nonumber \\
&- \un{\nabla}{}_{\ul{\alpha}}(\ov{s}\,\,^{\ul{\delta}}
\un{\nabla}{}_{\ul{\delta}}\tilde{c}_{\ul{\beta}})
+ \un{\nabla}{}_{\ul{\alpha}}\ov{s}_{\ul{\delta}}
\un{\nabla}^{\ul{\delta}}\tilde{c}_{\ul{\beta}}
+ O_{p-1}(R^{-3})\,.\label{ddc}
\end{align}
The third term on the right-hand side of~\eqref{ddc} is zero to
leading order because~$\tilde{c}_a$ only depends on angular
coordinates~\eqref{eqn:coordtransf}. Note that, to this order, whether
the dependence is on lower case angles or upper case ones is
irrelevant because of~\eqref{angles}. Also, from~\eqref{sup} we
have~$\un{\nabla}{}_a\ov{s}_b=\un{\nabla}{}_b\ov{s}_a$, so we get,
\begin{align}
\ov{s}_{\ul{\alpha}}\un{\nabla}{}_{\ul{\delta}}\un{\nabla}^{\ul{\delta}}
\tilde{c}_{\ul{\beta}} - \ov{s}{}^{\ul{\delta}}\un{\nabla}{}_{\ul{\delta}}
\un{\nabla}{}_{\ul{\alpha}}\tilde{c}_{\ul{\beta}} = \un{\nabla}{}_{\ul{\delta}}
(\ov{s}_{\ul{\alpha}}\un{\nabla}^{\ul{\delta}}\tilde{c}_{\ul{\beta}})
+ O_{p-1}(R^{-3})\,.\label{ddc2}
\end{align}
Plugging this into~\eqref{integrand1} yields,
\begin{align}
  \ov{\mathbb{U}}\,^{ab}\,_c\ov{n}_a \ov{s}_b X^c &=
  \un{\mathbb{U}}\,^{ab}\,_c\ov{n}_a \ov{s}_b X^c
  +\ov{n}_bX^b\un{\nabla}{}_a[\ov{s}{}^c
  \un{\nabla}{}^a\tilde{c}_c]\nonumber\\&+ \ov{n}{}^cX^b
  \un{\nabla}{}_a[\ov{s}_b\un{\nabla}^a\tilde{c}_c]
  + O_{p-1}(R^{-3})\,.\label{integrand2}
\end{align}
The vector $\p_{\ul{\alpha}}$ is a Killing vector with respect to the
upper case background metric~$B_{ab}$, so we can write,
\begin{align}
\Lie_X B_{ab} =2\ov{\nabla}_{(a}[B_{b)c}X^{c}]=0 \,. \label{Killingeq}
\end{align}
Naturally, we have that,
\begin{align}
	&(\p_{T})^a = -(B^{-1})^{ab}\ov{\nabla}_b T\,,\non\\
  &(\p_{\ul{i}})^a = (B^{-1})^{ab}\ov{\nabla}_b X^{\ul{i}}\,,
  \label{coordvectors}
\end{align}
which, together with \eqref{Killingeq}, gives
\begin{align}
	\ov{\nabla}_aX^b=0\,.\label{nabX}
\end{align}
Note that this result implies~$\un{\nabla}{}_aX^b=O_{p-1}(R^{-2})$,
because the difference between the background metrics has a
fall-off~\eqref{diffBb}, and this allows us to push $X$ through the
covariant derivative in \eqref{integrand2} while only getting higher
order additional terms. From~\eqref{diffBb} and~\eqref{alpha}, we can
easily see that,
\begin{align}
\ov{\alpha}=1+O_{p}(R^{-1})\,,\label{alphaexp}
\end{align}
and hence that,
\begin{align}
\un{\nabla}{}_{\ul{\alpha}}\ov{n}_{\ul{\beta}} =
O_{p-1}(R^{-2})\,,\label{deln}
\end{align}
Now, using~\eqref{nabX} and~\eqref{deln} in the
integrand~\eqref{integrand2}, we find,
\begin{align}
\ov{\mathbb{U}}\,^{ab}\,_c\ov{n}_a \ov{s}_b X^c
&= \un{\mathbb{U}}\,^{ab}\,_c\un{n}{}_a \un{s}{}_b X^c
+ \un{\nabla}{}_a[\ov{n}_bX^b\ov{s}{}^c
\un{\nabla}^a\tilde{c}_c] \nonumber\\
& + \un{\nabla}{}_a[\ov{n}{}^cX^b\ov{s}_b
\un{\nabla}^a\tilde{c}_c] + O_{p-1}(R^{-3})\,.\label{integrand3}
\end{align}
In the first term on the right-hand side we have replaced~$\ov{n}_a$
with~$\un{n}{}_a$ because, to leading order, they are
equal~\eqref{alphaexp}. Moreover, we have replaced~$\ov{s}_a$
with~$\un{s}{}_a$ because the antisymmetry of the first two indices
of~$\ov{\mathbb{U}}\,^{ab}\,_c$ guarantees that whichever component
of~$\ov{s}_a$ that is not orthogonal to~$\un{n}{}_a$ vanishes. The
last two terms are total divergences on the sphere and thus integrate
to zero. Finally, we get the result,
\begin{align}
H(X,\mathcal{S})=H(X,\mathcal{S}_{0})\,.\label{Hequal}
\end{align}
Let us now use this in order to find how the ADM momentum transforms
under~\eqref{eqn:coordtransf}:
\begin{align}
P^{\textrm{ADM}}_{\ul{\alpha}}:&=H(\p_{\ul{\alpha}},\mathcal{S}_{0})
=H(\p_{\ul{\alpha}},\mathcal{S})\nonumber\\
&=\Lambda_{\ul{\alpha}}{}^\alpha H(\p_{\alpha},\mathcal{S})
=:\Lambda_{\ul{\alpha}}{}^\alpha P^{\textrm{ADM}}_{\alpha}\,,\label{momtransffinal}
\end{align}
where in the second equality we used~\eqref{Hequal} and in the third
we used the fact that~$\Lambda_{\ul{\alpha}}{}^\alpha$ are constants
while the rest of the terms in the Jacobian are of order~$O(R^{-1})$,
so that they cannot contribute to the integral. This is the result
that we wanted~\eqref{eqn:ADMmom-transformation} and we state it
concisely in the following theorem:
\begin{thm}\label{Thm:T1}
Let~$(M,g)$ be an asymptotically flat spacetime and a solution of
Einstein's equations with stress-energy tensor components decaying
as~$O_0(R^{-4})$. Then, any permissible coordinate change transforms
the ADM 4-momentum as
\begin{align}
P^{\textrm{ADM}}_{\ul{\alpha}}=\Lambda_{\ul{\alpha}}{}^\alpha
P^{\textrm{ADM}}_{\alpha}\,.
\end{align}
\end{thm}

\paragraph*{Discussion:} The definition of asymptotic flatness at
spatial infinity given above makes no assumption about the linear
momentum, and nor should such a definition in general. By the result
on the ADM~$4$-momentum above~\eqref{momtransffinal} however, assuming
that the spacetime extends long enough near spatial infinity, we can
transform to an asymptotic rest-frame, or just rest-frame for short,
which we define as a slice in which the linear momentum vanishes. If
we wish, we can then refine the definition of asymptotic flatness
within this preferred slice. In view of the boost
theorem~\cite{ChrMur81}, we expect that given suitable initial data,
with appropriate care, our requirements on the asymptotics, to be
stated momentarily, can be propagated long enough in time to apply our
results.  We required that Einstein's equations are satisfied due to
the fact that~$P^{\textrm{ADM}}$ was defined according to the
corresponding action. That said, we expect that different actions
would yield similar results, but with different definitions of the
canonical stress tensor~\eqref{Tfalloff}.

\section{Conformal Flatness of Boosted Slices}
\label{section:Boosted-not-CF}

In this section, the Cotton-York tensor of the lower case spatial
metric is computed assuming the upper case slice to have zero ADM
linear momentum and the coordinate change to be given
by~\eqref{eqn:coordtransf}. For that a stronger definition of
asymptotic flatness is needed, namely, assumptions have to be made on
the first order terms in~$R^{-1}$ of the metric components. It turns
out that a crucial component of the Cotton-York tensor is given by the
boost vector itself. In the presence of linear momentum this component
gives the leading obstruction to conformal flatness in the lower case
foliation. Throughout this section we shall be concerned with
coordinate transformations of the form,
\begin{align}
    X^{\ul{\alpha}} =\Lambda^{\ul{\alpha}}{}_\alpha x^\alpha
  +c^{\ul{\alpha}}(\theta,\phi) +
  O_{4}(R^{-1})\,,\label{eqn:coordtransf2}
\end{align}
where we can take the error term in terms of the upper case radial
coordinate because of the equivalence of orders implied
by~\eqref{Rr2}.

\subsection{Strong asymptotic flatness}

We call a globally hyperbolic asymptotically flat spacetime with $p=3$
{\it strongly asymptotically flat of order~$O_3(R^{-2})$ at spatial
  infinity} if there exist coordinates~$X^{\ul{\mu}}=(T,X,Y,Z)$
defining a rest-frame in which, in a neighborhood of spatial infinity,
the spatial metric takes the form,
\begin{align}
\gamN_{\ul{ij}}&=\psi^4\big(\delta_{\ul{ij}}+h_{\ul{ij}}\big)\,,\label{strongaf1}
\end{align}
where~$h_{\ul{ij}}=O_3(R^{-2})$, and we fix the ambiguity in this
decomposition by taking~$\psi=1+\tfrac{m}{2R}$, whilst the lapse and
shift satisfy,
\begin{align}
A&=1-\frac{m}{R}+O_3(R^{-2})\,,&\,\,
B^{\ul{i}}&=O_3(R^{-2})\,.\label{strongaf2}
\end{align}
To highlight the differences between the notion of strong asymptotic
flatness and its weaker version given in
section~\ref{section:Asymptotic_flatness}, the former can be written
in a more concise way:
\begin{align}
  g_{\ul{\alpha\beta}}=\eta_{\ul{\alpha\beta}}
  +\frac{2m}{R}\delta_{\ul{\alpha\beta}}+O_3(R^{-2})\,.
\end{align}
Ultimately this amounts to requiring that the spacetime is
asymptotically flat with~$p=3$ and the coefficient of the~$R^{-1}$
term is~$2m\delta_{\ul{\alpha\beta}}$. We are not aware of a general
theorem guaranteeing that such fall-off will be propagated from
initial data, but this definition is satisfied by the Kerr-Newman
metric, and by the Schwarzschild metric with vanishing error terms,
and so is not absolutely prohibitive. It is similar in spirit but not
identical to the notion of strong asymptotic flatness employed
in~\cite{ChrKla93}, but we expect that we could adjust our definition
to match the conventions therein.

\subsection{Definition of conformal flatness}
\label{subsection:Obstruction}

It is well known that in three dimensions conformal flatness is
characterized by the vanishing of the Cotton, or equivalently
Cotton-York, tensor~\cite{Yor71,Yor72}. Working in the lower case
foliation, the Cotton tensor and the Cotton-York tensor associated
with~$\gamma_{ij}$ are given by,
\begin{align}
C_{abc}&:=D_{c}\left(R_{ab} -
\tfrac{1}{4}R\gamma_{ab}\right)-D_{b}\left(R_{ac} -
\tfrac{1}{4}R\gamma_{ac}\right)\nonumber\\
C^{ab}&:=-\frac{1}{2}\epsilon^{acd} C_{ecd}
\gamma^{eb} = \epsilon^{cd(a}D_cR^{b)}{}_d\,,\label{CYdef}
\end{align}
respectively, where the last equality makes use of the fact
that~$C^{ab}$ is a symmetric
tensor. Here,~$\epsilon^{bcd}:=n_a\epsilon^{abcd}$
and~$\epsilon^{abcd}$ is the Levi-Civita totally antisymmetric tensor
with indices raised with the metric $g_{ab}$. The definitions for the
upper case foliation are analogous. An important point to make here is
that if a metric is conformally flat, then there is a coordinate
system~$X^{\ul{\alpha}}$ in which we can write that locally,
\begin{align}
\gamN_{\ul{ij}} = \Omega^4 \delta_{\ul{ij}}\,.
\end{align}
We then say that~$\gamN_{ab}$ is {\it explicitly conformally flat} in
coordinates~$X^{\ul{\alpha}}$. Taking the spacetime to be strongly
asymptotically flat of order~$O_3(R^{-2})$, our primary assumption,
the upper case Cotton-York tensor is easily seen to be at worst,
\begin{align}
{}^{\textrm{\tiny{(N)}}}\!C^{\ul{ij}}=O(R^{-5})\,.
\end{align}
We call such a non vanishing Cotton-York tensor an upper case
obstruction to conformal flatness. Any such obstruction must be, in
some sense, generated by the traceless part of~$h_{\ul{ij}}$.

\subsection{Conformal flatness and the boost metric}
\label{subsection:Boost-CF}

Let us consider a spacetime which is strongly asymptotically flat of
order~$O_3(R^{-2})$. We want to show the result that the boost metric
components~\eqref{eqn:GmN->gmn} have the same type of fall-off near
spatial infinity as the upper case spatial metric components in an
appropriate set of spatial coordinates. This observation will be
helpful when computing the lower case Cotton-York tensor. The lower
case metric can be written as,
\begin{align}
\gamma_{ij} = &g_{\ul{\alpha\beta}}J^{\ul{\alpha}}{}_i J^{\ul{\beta}}{}_j\nonumber\\
= &\gamN_{\ul{ij}}\phi^{\ul{i}}{}_i\phi^{\ul{j}}{}_j
- W^2v_iv_j+A^{-2}B_{\ul{i}}B^{\ul{i}}W^2v_iv_j \nonumber\\
&-A^{-1}B_{\ul{i}}\phi^{\ul{i}}{}_{(i}Wv_{j)}\,,\label{gam=gJJ}
\end{align}
where, in the second equality, we have used~\eqref{eqn:Jacobian}. Then
the boost metric is exactly,
\begin{align}
\gB_{ij}=\gamN_{\ul{ij}}\phi^{\ul{i}}{}_i\phi^{\ul{j}}{}_j
+ A^{-2}B_{\ul{i}}B^{\ul{i}}W^2v_iv_j -A^{-1}B_{\ul{i}}\phi^{\ul{i}}{}_{(i}Wv_{j)}\,.
\end{align}
Strong asymptotic flatness on our metric~$g_{ab}$ gives,
\begin{align}
\gB_{ij} = \gamN_{\ul{ij}}\phi^{\ul{i}}{}_i\phi^{\ul{j}}{}_j +
O_3(R^{-2})\,,\label{gBaflat}
\end{align}
which does not depend on~$J^{T}{}_i$, so there must be a set of
spatial coordinates~$x^{\hat{i}}$ that allows us to write
\begin{align}
  \gB_{\hat{i}\hat{j}}=\psi^4(\delta_{\hat{i}\hat{j}}+\mathbbmss{h}_{\hat{i}\hat{j}})\,,
\end{align}
with~$\mathbbmss{h}_{\hat{i}\hat{j}}=O_3(R^{-2})$. Notice that having
made no assumption on the form of the boost,~$\gB_{ij}$ inherits the
asymptotic form of the upper case metric~\eqref{strongaf1}. In fact,
these coordinates are easily seen to be given
by~$x^{\hat{i}}=X^{\ul{i}}$, so that the full composite transformation
is given by
\begin{align}
\hat{t}&=t=\bar{W}(T+\bar{v}_{i}\delta^i{}_{\ul{i}}X^{\ul{i}}) + C^t(\Theta,\Phi) +
O_{4}(R^{-1})\,,\nonumber\\ x^{\hat{i}}&=X^{\ul{i}}\,,\label{mixedcoord}
\end{align}
which renders the spatial part of the
Jacobian~$\phi^{\ul{i}}{}_{\hat{i}}=\delta^{\ul{i}}{}_{\hat{i}}$. Note
that this coordinate transformation does not give a Lorentz
transformation at leading order, and hence must be treated carefully
when evaluating asymptotic charges. Although the slice is boosted, the
time derivative associated with these coordinates still coincides
with~$\p_T$, which means that the solution {\it still} appears time
independent at order~$O(R^{-1})$ in the transformed tensor basis.

It is interesting to note also that in the static case, taking the
upper case coordinates to have vanishing shift, the error term
in~\eqref{gBaflat} vanishes and~$\gB_{ab}$ is conformally flat
whenever the upper case spatial metric is. Moreover, Einstein's
equations were not used to reach this result, meaning that it is fair
to say that the following fact is purely geometrical: \textit{in a
  static spacetime with a foliation with vanishing shift in which the
  spatial metric is conformally flat, the boost metric relative to
  that foliation is conformally flat with the same conformal
  factor}. More generally, since the boost metric is conformally
related to~$\delta_{\hat{i}\hat{j}}+\mathbbmss{h}_{\hat{i}\hat{j}}$,
we can say that the obstruction to conformal flatness in the boost
Cotton-York tensor is at worst~$O(R^{-5})$. Naturally, we recover the
precise obstruction of the upper case Cotton-York tensor continuously
as~$v\to0$. In other words in strongly asymptotically flat spacetimes
of order~$O_3(R^{-2})$, boost metrics have the same obstruction to
conformal flatness as the spatial metric in the preferred
rest-frame. On this basis one would therefore expect that the spatial
metric in such a boosted slice would pick up an obstruction to
conformal flatness at lower order in~$R^{-1}$. This we examine in the
following.

\subsection{The lower case Cotton-York tensor}
\label{subsection:Cotton-York}

From~\eqref{momtransffinal} we can see that if we assume the upper
case slice to have zero ADM linear momentum, then any slice that we
get by changing coordinates according to~\eqref{eqn:coordtransf2} has
non-vanishing linear momentum if and only if~$\bar{v}_i\neq 0$. In the
last section we saw that the boost metric of Schwarzschild spacetime
is conformally flat. Then, looking at~\eqref{eqn:GmN->gmn}, we expect
that~$\gamma_{ij}$ is not. In this section we compute the lower case
Cotton-York tensor using~\eqref{gam=gJJ} to show that our expectations
are correct for a large class of spacetimes. We begin by assuming that
our spacetime is strongly asymptotically flat of
order~$O_3(R^{-2})$. While it is possible to do this computation
directly, it proves more efficient to use the conformal invariance of
the Cotton tensor and compute it for a metric that is conformal to the
lower case spatial metric. For that, let us expand~$Wv_i$
under~\eqref{eqn:coordtransf2},
\begin{align}
Wv_i=A(\bar{W}\bar{v}_i-\p_i c^T)+O_{3}(R^{-2})\,,\label{wvexp}
\end{align}
and plug it in to~\eqref{gam=gJJ} to get,
\begin{align}
\tilde{\gamma}_{ij}&:=\psi^{-4}\gamma_{ij}\nonumber\\
&=\delta_{ij}+\frac{4m}{R}\bar{W}^2\bar{v}_i\bar{v}_j
+2\bar{W}^2\bar{v}_{(i}\p_{j)}c^T\nonumber\\
&\quad+ 2\phi_{\ul{i}(i}\p_{j)}c^{\ul{i}}+O_{3}(R^{-2})\,,
\end{align}
where the second equality is obtained from equations~\eqref{jac2}
and~\eqref{wvexp}
and~$\phi_{\ul{i}i}:=\delta_{\ul{ij}}\phi^{\ul{j}}{}_{i}$. The
Levi-Civita connection associated with~$\tilde{\gamma}_{ij}$ is,
\begin{align}
\tilde{\Gamma}_i{}^k{}_j=&\frac{2m}{R^2}\bar{W}(\bar{v}_i\bar{v}_js^k
+\bar{v}_i\bar{v}^ks_j - \bar{v}_j\bar{v}^ks_i)\nonumber\\
&+\bar{W}\bar{v}^k\p_i\p_jc^T+\phi_{\ul{i}}{}^k\p_i\p_jc^{\ul{i}}
+O_{2}(R^{-3})\,,\label{confLeviCiv}
\end{align}
where~$\phi_{\ul{i}}{}^i:=\delta^{ij}\phi_{\ul{i}}{}_{j}$ and~$s_i$
defined as,
\begin{align}
s_i := L\p_iR\nonumber\,,\quad L^{-2}:=\gBi^{ij}\p_i R\p_j R\,.
\end{align}
Here we raise and lower the indices on~$s_i$ and~$\bar{v}_i$
with~$\gamma^{ij}$. In order to compute the Ricci tensor of the
conformal metric, we will need to take one derivative of
$s_i$. From~\eqref{Rr1}, we get,
\begin{align}
\p_is_j=\frac{1}{R}(\gB_{ij}-s_is_j)+O_{2}(R^{-2})\,,
\end{align}
and hence,
\begin{align}
\tilde{R}_{ij}=&\p_k\tilde{\Gamma}_i{}^k{}_j-\p_j\tilde{\Gamma}_i{}^k{}_k
+\tilde{\Gamma}_i{}^l{}_j\tilde{\Gamma}_k{}^k{}_l-\tilde{\Gamma}_i{}^l{}_k
\tilde{\Gamma}_j{}^k{}_l\nonumber\\
=&-\frac{2m}{R^{3}}\bar{W}^2(12\bar{v}_{[i}s_{k]}\bar{v}_{[j}s_{l]}\delta^{kl}
-\bar{v}_i\bar{v}_j-\bar{v}_k\bar{v}^k\delta_{ij})\nonumber\\
&+O_1(R^{-4})\,.
\end{align}
Note that, to leading order, the Ricci tensor of the conformal metric
does not depend on supertranslations. We can finally compute the
Cotton-York tensor of the conformal metric using~\eqref{CYdef},
\begin{align}
\tilde{C}^{ij}=&\frac{30\bar{W}^2m}{R^4}s_k\bar{v}_l
\epsilon^{kl(i}\left\{\frac{1}{5}\bar{v}^{j)}+\bar{v}^{j)}[\bar{v}_ms^m]^2
-s^{j)}\bar{v}_ms^m\right\}\nonumber\\
&+O_0(R^{-5})\,,\label{lcCotton41}
\end{align}
In order to obtain the lower case Cotton-York tensor from the
conformal one, we use the conformal invariance of the Cotton tensor
and verify that, to leading order, the Cotton-York tensors must agree,
\begin{align}
C_{ijk}=\tilde{C}_{ijk}\Leftrightarrow C^{ij}=\tilde{C}^{ij}
+O(R^{-5})\,.
\end{align}
Notice that $\bar{v}_i$ and $s_i$ cannot be parallel because, in
Cartesian coordinates, $\bar{v}_i$ is constant. To leading order, all
the five independent components of~$C^{ij}$ (symmetric and trace-free)
vanish if and only if~$\bar{v}=0$ or~$m=0$, except~$C^{ij}s_is_j$
which is zero regardless of the values of the constants. This implies
that, if we assume our metric to have a `rest-frame'
($P^{\textrm{ADM}}_{\ul{i}}=0$) that is close to conformal flatness in
the asymptotic sense of~\eqref{strongaf1}-\eqref{strongaf2}, no slice
with~$\bar{v}_i\neq0$ can be conformally flat. It is interesting to
note that this fact is purely geometrical, in the sense that it does
not assume GR to hold. It is only when we talk about linear momentum
that this ceases too be true, because its definition and
transformation law~\eqref{eqn:ADMmom-transformation} rely on
GR. However we do expect that similar results can be obtained for
different theories. For clarity we state this result in the following
theorem:

\begin{thm}\label{Thm:T2}
    Let~$(M,g)$ be a strongly asymptotically flat spacetime of
    order~$O_3(R^{-2})$ at spatial infinity and a solution of
    Einstein's equations with non-trivial~$m$ and stress-energy tensor
    components decaying as $O_0(R^{-4})$. Then, there is no
    permissible slice with non-zero ADM linear momentum which is
    conformally flat.
\end{thm}

\subsection{The Kerr case}
\label{subsection:Kerr}

It is straightforward to see that the Kerr spacetime satisfies the
hypotheses of both Theorems~\ref{Thm:T1} and~\ref{Thm:T2}. Therefore
there can be no conformally flat boosted slice of Kerr. In fact it is
already known~\cite{GarPri00,Kro03,Kro04} that there is no such slice
with vanishing linear momentum in Kerr either. Presently, as a
corollary of Theorem~\ref{Thm:T2}, we recover the latter result in the
special case that the slice is axially symmetric. Details of the
calculations of this section can be found the Mathematica notebook
that accompanies the paper~\cite{DuaHil19_web}.

We start with Boyer-Lindquist coordinates and adjust the radial
coordinate~$R_{\textrm{BL}}$ as,
\begin{align}
R_{\textrm{BL}}&=\psi^2R\,,
\end{align}
with~$\psi$ defined as before.  Constructing Cartesian coordinates in
the standard way from~$(T,R,\Theta,\Phi)$ brings the metric into the
form~\eqref{strongaf1} employed in Theorem~\ref{Thm:T2}. Computing the
Cotton-York tensor, one readily finds an obstruction to conformal
flatness or order~$O(R^{-7})$ at large radius. Therefore our aim would
be to adjust the slice so that this obstruction is somehow
absorbed. We consider only axisymmetric slices, and so make the
ansatz,
\begin{align}
  t&= \bar{W}(T+\bar{v}_{i}\delta^i{}_{\ul{i}}X^{\ul{i}})
  +C_{(0)}^t+R^{-1}C_{(1)}^t\,,
\end{align}
with~$C_{(0)}^t$ and~$C_{(1)}^t$ functions of~$\Theta$ to be
determined. Working with axisymmetric slices means that we end up with
only a simple ODE analysis to perform. Generalizing this would instead
require treating a PDE problem. Adding higher order terms to this
ansatz will not affect the calculations to the order at which we
work. Presently we do not alter the spatial coordinates, since doing
so will only complicate the computation, and can not help to impose
conformal flatness on the adjusted spatial slice, which is determined
solely by the choice of~$t$. By Theorem~\ref{Thm:T2} we must
furthermore choose~$\bar{v}_{\ul{i}}$ trivial, otherwise there will be
an obstruction to conformal flatness of order~$O(R^{-4})$ on the
adjusted slice. Computing the Cotton tensor of the lower case spatial
metric in powers of~$R^{-1}$ reveals that there is an obstruction to
conformal flatness at order~$O(R^{-5})$ unless,
\begin{align}
C^t_{(0)}&=c_1+c_2\cos^2\Theta\,,
\end{align}
with~$c_1$ and~$c_2$ arbitrary real constants. There is furthermore an
obstruction of order~$O(R^{-6})$ unless~$c_2=0$; in other words the
supertranslation term must belong to the Poincar\'e class. Using these
conditions and computing one order further we find that there is no
choice of~$C_{(1)}^t$ that removes the~$O(R^{-7})$ obstruction. In
particular, we must have,
\begin{align}
  \sin\Theta\,\p^2_\Theta C^t_{{(1)}} +3\cos\Theta\,\p_\Theta
  C^t_{{(1)}}=0\,,
\end{align}
but that even when this condition is satisfied there remains an
obstruction at the same order. Thus the Kerr spacetime admits no
axially symmetric conformally flat spatial slice.

\section{Conclusions}\label{section:Conclusions}

Working with asymptotically flat spacetimes and using the DF formalism
we have made a number of interesting findings. Starting from a set of
coordinates in which the metric has good asymptotic behavior and
performing a boost that preserves this fall-off of the metric near
spatial infinity, we first found that the ADM $4$-momentum is governed
solely by the leading Lorentz transformation of the boost even in the
presence of supertranslation terms, generalizing the result beyond the
Poincar\'e group.

We then restricted our notion of asymptotic flatness in order to study
conformal flatness of boosted frames. The special property of our
class is that there exist rest-frames, slices with vanishing linear
momentum, in which the spatial metric is close to conformal
flatness. Working with spatial slices that can be boosted with respect
to such a rest-frame we showed that the boost metric inherits
properties from its unprojected counterpart. Using this fact and
restricting our attention to boosted slices with nonvanishing linear
momentum, from which it follows that the ADM mass and asymptotic boost
must be nontrivial by our first result, we found that the Cotton
tensor in the boosted slice picks up an~$O(R^{-4})$ term. Linear
momentum therefore serves as an obstruction to conformal flatness in
these spacetimes.

Turning our attention to the Kerr spacetime we recovered a special
case of the result~\cite{GarPri00,Kro03,Kro04} that axisymmetric
slices in this spacetime can not be conformally flat. More generally
it is clear that even in strongly asymptotically flat spacetimes of
order~$O_3(R^{-2})$, adjustment of slices can only annihilate an
obstruction to conformal flatness if that obstruction has a very
special structure. A complete characterization of that structure is
still lacking, however.

From a practical point of view, for applications in numerical
relativity, our findings suggest that it may be natural to adopt a
conformally flat boost metric as an ingredient in the construction of
initial data. For that one could employ a method similar to the
standard conformal-transverse-traceless decomposition of the
constraints. Such a construction would then proceed in the spirit
of~\cite{RucHeaLouZlo14,HeaRucLou15}. Likewise a natural suggestion
for the extrinsic curvature, which still needs to be properly
formalized, would be to make it `essentially' a Lie-derivative of
the 3-metric along the boost vector. In the case of a single black
hole, such data would reduce to a boosted slice of
Schwarzschild. Therefore we expect that data so constructed would
contain less junk-radiation as compared with the present
moving-puncture approach. These physically motivated choices do not
obviously lead to a mathematically simple formulation of the
constraints, so we postpone further discussion for future work.

\acknowledgments

We are grateful to Bernd Br\"ugmann, Edgar Gasperin, Mark Hannam,
Jos\'e Nat\'ario and Juan Valiente Kroon for helpful discussions
and/or comments on the manuscript. This work was supported in part by
the FCT (Portugal) IF Program~IF/00577/2015, PTDC/MAT-APL/30043/2017
and PD/BD/135511/2018. DH also gratefully acknowledges support offered
by IUCAA, Pune, where part of this work was completed.

\appendix

\section{Propagation of asymptotic flatness from initial data}
\label{section:App_Sob}

It has been shown that requiring initial data for vacuum GR to
be asymptotically flat, for some definition of the term, gives a time
development that preserves the asymptotic fall off~\cite{ChrMur81}. As
our definition of asymptotic flatness is a set of conditions on the
whole of a boost region, it is interesting to check whether this
definition is a consequence of the initial data requirements
of~\cite{ChrMur81}. If that is true, then we need only to impose
conditions on an initial slice that are sufficient to guarantee that
they are preserved in a boost region. Let~$U$ be any open set
in~$\mathbb{R}^3$ and let~$\sigma$ be the function,
\begin{align}
    \sigma(R)=(1+R^2)^{1/2}\,.
\end{align}
The weighted Sobolev space~$H^k_{\delta}(U)$, with $s\in \mathbb{N}$
and $\delta\in\mathbb{R}$, is the class of all functions $u$ on $U$
with values in some finite dimensional vector space $V$, defined by
the norm:
\begin{align}
    ||u||_{H^k_{\delta}(U)}:=\sum_{j=0}^{k}||\sigma^{\delta+j}D^ju||_{L^2(U)}\,.
\end{align}
The first statement of the boost theorem, as stated in
\cite{ChrMur81}, is the following: Let $\gamN$ be a Riemannian metric
and $\KN$ a 2-covariant symmetric tensor field on $\Sigma_t$. If,
\begin{align}
  \gamN-e\in H^k_{\delta+\frac{1}{2}}(\Sigma_t)\quad,\quad \KN
  \in H^{k-1}_{\delta+\frac{3}{2}}(\Sigma_t)\,,
\end{align}
where $k\geqslant 4$, $\delta>-2$ and $e$ is the 3-dimensional flat
metric, then there exists a metric $g$ solution of Einstein's
equations in a boost-type domain $\Omega$, such that $g-\eta\in
H^k_{\delta}(\Omega)$ and $(\gamN,\KN)$ are respectively the first and
second fundamental forms of $g$ associated with $\Sigma_t$. On
$\Omega$ we now define the function:
\begin{align}
    \tau(t,R):=\frac{t}{\sigma(R)}\,,
\end{align}
whose level surfaces define a foliation,
\begin{align}
  \Omega=\bigcup_{\tau\in I_\theta}\Sigma_\tau\quad,\quad I_\theta
  =(-\theta,\theta)\,.
\end{align}
Then Lemma 2.4 in~\cite{ChrMur81} states that, for each $\tau\in
I_\theta$, the following inclusion holds and is continuous:
\begin{align}
  H^k_{\delta}(\Omega)\subset
  H^{k-1}_{\delta+\frac{1}{2}}(\Sigma_\tau,\Omega)\,, \label{restriction}
\end{align}
where the space $H^k_{\delta}(\Sigma_\tau,\Omega)$ is defined by the
norm,
\begin{align}
  ||u||_{H_{k,\delta}(\Sigma_\tau,\Omega)}^2
  :=\sum_{j=0}^{k}||D_t^ju|_{\Sigma_\tau}||_{H^{k-j}_{\delta+j}(\mathbb{R}^n)}\,.
\end{align}
\eqref{restriction} then gives that, for each $j\leqslant k$,
\begin{align}
    D_t^j(g-\eta)|_{\Sigma_\tau}\in H^{k-1-j}_{\delta+\frac{1}{2}+j}(\mathbb{R}^n)\,,
\end{align}
where $D_t$ is a time derivative. By definition of the weighted
Sobolev norm we know that if we take a spatial derivative $\bar{D}$,
we get,
\begin{align}
  \bar{D}^iD_t^j(g-\eta)|_{\Sigma_\tau}\in
  H^{k-1-i-j}_{\delta+\frac{1}{2}+i+j}(\mathbb{R}^n)\,,\label{ddu1}
\end{align}
where $i+j$ is the number of derivatives taken in all directions. We
introduce the weighted Sobolev norms defined by,
\begin{align}
  &||u||_{W^{k,\infty}_{\delta}(U)}:=\sum_{j=0}^{k}\textrm{ess}
  \,\textrm{sup}_{U}(|\sigma^{\delta+j}\bar{D}^ju|)\,,\\
  &||u||_{W^{k,\infty}_{\delta}(\Sigma_\tau,\Omega)}^2
  :=\sum_{j=0}^{k}||D_t^ju|_{\Sigma_\tau}||_{W^{k-j,\infty}_{\delta+j}(\mathbb{R}^n)}
\,,\label{funky1}
\end{align}
where~\eqref{funky1} can be written in a more convenient way as,
\begin{align}
  ||u||_{W^{k,\infty}_{\delta}(\Sigma_\tau,\Omega)}^2
  &= \sum_{j=0}^{k}\sum_{i=0}^{k-j}\textrm{ess}\,
  \textrm{sup}_{\mathbb{R}^n} (|\sigma^{\delta+i+j}
  \bar{D}^iD^j_tu|_{\Sigma_\tau}|)\nonumber\\
  &= \sum_{j=0}^{k}\sum_{i=0}^{k-j}
  ||\bar{D}^iD^j_tu|_{\Sigma_\tau}||_{W^{0,\infty}_{\delta+i+j}(\mathbb{R}^n)}
\,.\label{funky2}
\end{align}
In~\cite{Bar86}, equation (1.9) shows the Sobolev embedding result
that we need,
\begin{align}
  H^k_{\delta+\frac{3}{2}}(\mathbb{R}^n) \subset
  W^{0,\infty}_{\delta}(\mathbb{R}^n)\,,\label{bartnikembed}
\end{align}
for any~$k\geqslant 2$. Note that this result seems different from the
one in~\cite{Bar86} because our definitions for the weighted Sobolev
norms are more in line with~\cite{ChrMur81}, where~$\delta$ is defined
differently. From~\eqref{ddu1} and~\eqref{bartnikembed} we find,
\begin{align}
  \bar{D}^iD_t^j(g-\eta)|_{\Sigma_\tau}\in W^{0,\infty}_{\delta-1+i+j}
  (\mathbb{R}^n)\,,\label{ddu2}
\end{align}
with~$k\geqslant i+j+3$. Then, if we want that~$p$ derivatives in any
directions improve the fall off of the metric, we must
choose~$k\geqslant p+3$. This, together with~\eqref{funky2} implies
that,
\begin{align}
    g-\eta\in W^{p,\infty}_{\delta-1}(\Sigma_\tau,\Omega)\,,\label{getafunky}
\end{align}
which in turn implies our definition of asymptotic flatness,
\begin{align}
	g_{\ul{\alpha\beta}}=\eta_{\ul{\alpha\beta}}+O_{p}(R^{-1})\,,
\end{align}
with~$p\geqslant 1$, given that we choose~$\delta=2$. Note that the
possible choices of~$k$ and~$\delta$ that give the desired asymptotic
conditions trivially satisfy the requirements of the boost theorem. It
is thus shown that our definition of asymptotic flatness holds if we
require our initial data to have the asymptotic behavior of the boost
theorem.

\bibliographystyle{unsrt}
\bibliography{CF_DF.bbl}{}

\end{document}